\documentclass[12pt]{article}
\usepackage{epsfig,amsfonts,amssymb}
\usepackage{hyperref}
\usepackage{cite}
\input epsf.sty
\usepackage{cite}

\def\ZZZ{{\hbox{ Z\kern-1.6mm Z}}}
\def\RRR{{\hbox{ R\kern-2.4mm R}}}
\def\CCC{{\hbox{ C\kern-2.0mm C}}}
\def\zzz{{\hbox{z\kern-1mm z}}}

\newcommand{\vt}{\vartheta}

\newcommand{\qeq}{{\hbox{=\kern-2.3mm ? \kern.5mm }}}
\renewcommand{\qeq}{=}

\newcommand{\II}{{\cal I}}
\newcommand{\AAA}{{\cal A}}

\newcommand{\FF}{{\cal F}}

\newcommand{\MM}{{\cal M}}

\newcommand{\OO}{{\cal O}}

\newcommand{\wt}{\widetilde}
\newcommand{\wh}{\widehat}

\newcommand{\NN}{{\cal N}}

\newcommand{\hI}{\wh\II}

\newcommand{\be}{\begin{equation}}
\newcommand{\ee}{\end{equation}}
\newcommand{\ben}{\begin{eqnarray}\displaystyle}
\newcommand{\een}{\end{eqnarray}}

\newcommand{\bea}[1]{\begin{eqnarray}\label{#1} }
\newcommand{\eea}{\end{eqnarray}}

\newcommand{\refb}[1]{(\ref{#1})}

\newcommand{\sectiono}[1]{\section{#1}\setcounter{equation}{0}}

\def\one{{\hbox{ 1\kern-.8mm l}}}
\def\zero{{\hbox{ 0\kern-1.5mm 0}}}

\begin{document}

\baselineskip 24pt

\begin{center}
{\Large \bf
A Twist in the Dyon Partition Function\footnote{Dedicated to the
memory of Alok Kumar and Jaydeep Majumder.}}

\end{center}

\vskip .6cm
\medskip

\vspace*{4.0ex}

\baselineskip=18pt

\centerline{\large \rm   Ashoke Sen }

\vspace*{4.0ex}

\centerline{\large \it Harish-Chandra Research Institute}
\centerline{\large \it  Chhatnag Road, Jhusi,
Allahabad 211019, India}
\centerline{and}
\centerline{\large \it 
LPTHE, Universite Pierre et Marie Curie, Paris 6}
\centerline{\large \it 
4 Place Jussieu,  75252 Paris Cedex 05, France}

\vspace*{1.0ex}
\centerline{E-mail:  sen@mri.ernet.in, ashokesen1999@gmail.com}

\vspace*{5.0ex}

\centerline{\bf Abstract} \bigskip

In four dimensional string
theories with $\NN=4$ and $\NN=8$
supersymmetries one can often define
twisted index in a subspace of the moduli space which captures
additional information on the partition function than the ones contained
in the usual helicity trace index. 
We compute several such indices in type IIB string theory on
$K3\times T^2$ and $T^6$, and find that they 
share many properties with the usual helicity
trace index that captures the spectrum of quarter BPS states in
$\NN=4$ supersymmetric string theories.
In particular
the partition function is a modular form of a subgroup of
$Sp(2;\ZZZ)$  and the 
jumps across the walls of marginal stability are
controlled by the residues at the poles of the partition function. 
However for large charges the logarithm of this index
grows as $1/N$ times the entropy of a black hole carrying the same
charges where $N$ is the order of the symmetry generator
that is used to define the twisted index. We provide a macroscopic
explanation of this phenomenon using quantum entropy function
formalism. The leading saddle point corresponding to the attractor
geometry fails to contribute to the twisted index, but a $\ZZZ_N$
orbifold of the attractor geometry produces the desired
contribution.

\vfill \eject

\baselineskip=18pt

\tableofcontents

\sectiono{Introduction and Summary}

We now have a good understanding of the spectrum of dyons
in $\NN=4$ and $\NN=8$ supersymmetric string 
theories\cite{9607026,0412287,0505094,
0506249,0508174,0510147,0602254,
0603066,0605210,0607155,0609109,0612011,0702141,
0702150,0705.1433,0705.3874,0706.2363,
0708.1270,0802.0544,0802.1556,0803.2692,0806.2337,0807.4451,
0809.4258,0808.1746,0901.1758,
0907.1410,0911.0586,9903163,0506151,0506228,0803.1014,
0804.0651}. 
For large charges
the result for the degeneracy agrees with the macroscopic entropy
of a black hole carrying the same charges. 
We also have a good understanding of how to systematically
compute higher derivative 
corrections\cite{9307038,9812082,9906094,0506176,0506177} and 
quantum corrections\cite{0005003,0607138,0809.3304,0903.1477}
to the black hole entropy. 
Some of these corrections have already been used to test the
correspondence between the microscopic and black hole
entropies beyond the leading order.
Eventually
one hopes to be able to systematically 
compute the corrections to the black hole entropy
using these techniques, and compare the
results with the microscopic answer, thereby testing the
correspondence between macroscopic and microscopic entropies
to much finer detail. Some attempt to generalize these results to
half BPS black holes in the $\NN=2$ supersymmetric STU model
has also been made in \cite{0711.1971,0810.1233}.

On the microscopic side one often computes an index rather
than the absolute degeneracy, defined so that it receives contribution
only from the BPS states in the spectrum.
As a result these indices are protected and do not vary
continuously as we vary the moduli of the theory. In four
dimensions the standard index is the helicity trace index
$B_{2n}$ defined as follows\cite{9708062,9708130}
\be \label{edefb2n}
B_{2n} = {1\over (2n)!}\,
Tr\left[ (-1)^{2h} (2h)^{2n}\right] \, ,
\ee
where $h$ is the third component of the angular momentum of
a state in the rest frame, and the trace is taken over all states
carrying a given set of charges. In order that a given state gives a
non-vanishing contribution to this index, the number of supersymmetries
broken by the state must be less than or equal to $4n$. This is due to
the fact that for every pair of
broken supersymmetries we have a pair
of fermion zero modes whose quantization gives a bose-fermi
degenerate pair of states. As a result $Tr(-1)^{2h}$ will vanish unless
we insert a factor of $2h$ which prevents the cancelation between
these pair of states, thereby effectively soaking up the pair of fermion
zero modes. Thus if we have more than $4n$ broken supersymmetries,
and hence more that $4n$ fermion zero modes, then $B_{2n}$ does
not contain enough insertions of $2h$ to soak up all the fermion zero
modes, and the contribution to the trace
from such states vanishes. On the other hand if we have
states with
precisely $4n$ broken supersymmetries then $B_{2n}$ receives
contribution from these states, but not from any other states with
more than $4n$ broken supersymmetries. This makes $B_{2n}$
the ideal index for capturing protected information on states with
$4n$ broken supersymmetries. Some standard examples are 
$B_2$ for half BPS states in $\NN=2$ supersymmetric theories,
$B_4$ for half BPS states in $\NN=4$ supersymmetric theories,
$B_6$ for quarter BPS states in $\NN=4$ supersymmetric theories,
$B_{14}$ for 1/8 BPS states in $\NN=8$ supersymmetric theories
etc. The normalization in the definition of 
$B_{2n}$ has been adjusted
so that the contribution to the trace from the $4n$ fermion zero modes
due to broken supersymmetries exactly cancels the denominator
factor of $(2n)!$ except for a sign given by $(-1)^n$.

Given that on the macroscopic side the black hole entropy always
gives the absolute degeneracy whereas on the microscopic side we
compute the helicity trace index, one might wonder whether comparing the
two is justified. A resolution of this issue was proposed in
\cite{0903.1477} where it was shown how using the expression for the
degeneracy on the macroscopic side one can compute the
helicity trace index. This can then be compared with the
microscopic results. This argument will be reviewed in
\S\ref{ss6}.

In this paper we shall study a modified index obtained by
twisting the helicity trace index by an appropriate discrete
symmetry transformation -- both on the microscopic and the macroscopic
side -- and compare the results. For this we need to
restrict  the moduli to be on special subspaces of the moduli
space where the theory admits extra discrete symmetries generated
by an element $g$, 
and restrict the charges carried by the
dyon to be such that they are invariant under $g$. 
In this case we can
define a new twisted index as\footnote{Black holes carrying discrete
gauge charges have been discussed earlier in \cite{wil1,wil2}.}
\be \label{etw1}
B^g_{2n}\equiv {1\over (2n)!}\,
Tr\left[ g\, (-1)^{2h} (2h)^{2n}\right] \, .
\ee
If all the supersymmetry generators of the theory are invariant under
$g$ then our earlier counting holds and we conclude that
$B^g_{2n}$ does not receive any contribution from states which
breaks more than $4n$ supercharges. However suppose that
some of the broken supersymmetry generators are not
invariant under $g$. 
In that case the corresponding fermion zero modes are also
not invariant under $g$, and the contribution from these
modes in $Tr(-1)^{2h} g$ will not vanish.
As a result
we do not need any factor of $2h$ to soak up this pair of
fermion zero modes. If on the other hand we have a pair
of fermion zero modes which are invariant under $g$ then we need
a factor of $2h$ in the trace to soak up these zero modes.
{\it Thus $B^g_{2n}$ receives non-vanishing
contribution from states which break less than or equal to $4n$
$g$-invariant supersymmetries, -- the total number of broken
supersymmetries can be more than $4n$.} Conversely, if we have a
state that breaks certain number of supersymmetres, and if $4n$ of
these broken
supersymmetries are invariant under $g$, then the ideal $g$-twisted
index for capturing information about the spectrum of these states
is $B^g_{2n}$.
Since we expect $B^g_{2n}$ to have properties similar to that for
$B_{2n}$ ({\it e.g.} for wall 
crossing\cite{0005049,0010222,0101135,0206072,
0304094,0702146}), 
this allows us to introduce
an index in extended supersymmetric theories which behaves as an
index in a theory with less number of 
supersymmetries.\footnote{The use of such indices is not
new. In particular such an index has been analyzed in detail in
$\NN=4$ supersymmetric gauge theories
where it was found that
supersymmetric multi-monopole solutions preserving
quarter of the supersymmetries exist only on a subspace
of the moduli 
space\cite{0005275,0609055,9804160,0802.0761}. 
As a result
these do not contribute to the usual helicity trace index
$B_6$ which should have the same value everywhere in the
moduli space (leaving aside possible jumps across walls of
marginal stability). Nevertheless one can define appropriate index
to capture information about the multi-monopole states in the 
particular subspace of the moduli space where the solution 
exists.}

The main goal of this paper will be to compute such twisted
indices in type II string theory compactified on $\MM\times T^2$
where $\MM$ can be either $T^4$ or $K3$.
We choose $g$ to be the generator of 
a geometric $\ZZZ_N$ symmetry acting on
$\MM$ preserving 16 supersymmetries. For type II string
theory on $T^4\times T^2$ this requires $g$ to commute with 16 out
of the 32 unbroken supersymmetries, while for type IIB string theory
on $K3\times T^2$ this will require $g$ to commute with all the 16
unbroken supersymmetries. 
Examples of such $\ZZZ_N$ transformations
have been discussed in \cite{9508144,9508154}, and the dyon
spectrum on orbifolds of the original theory by these symmetries
(accompanied by a translation along a circle) have been analyzed
in \cite{0605210,0607155,0609109}. 
These (without the translations along the circle)
will be the $\ZZZ_N$ transformations we shall be using in
our analysis.
However 
here 
we do not
take the orbifold of the original theory; we simply use $g$
for defining a twisted partition function in the original theory.
In this theory we consider dyonic states preserving 4 supersymmetries
all of which are $g$ invariant. 
Such a dyon breaks 12 of the 16 $g$-invariant supersymmetries,
and
the relevant $g$-twisted index is $B^g_6$. 
We find explicit expression for 
this index in the examples described above, and find that
the index has properties similar to $B_6$ in $\NN=4$ supersymmetric
string theories, -- the helicity trace index used for
encoding information on $1/4$ BPS states in this theory.
In particular:
\begin{enumerate}
\item The index is given by the Fourier transform of the
inverse of a modular form of a subgroup of $Sp(2,\ZZZ)$.
\item The value of the index in different domains separated
by the walls of marginal stability is controlled by the same
partition function -- with the information on the domain being
encoded in the choice of contour along which the Fourier integral
is to be performed.
\item The jumps across the walls of marginal stability are controlled
by the residues at certain poles of the partition function. The resulting
expression for the jump follows the same wall crossing formula as the
usual $B_6$ index.
\item The growth of the index for large charges is controlled by
another set of poles of the partition function. 
\end{enumerate}
There is however an important difference. If we associate an
`entropy' to this index defined by taking its logarithm, we find that
for large charges the entropy is given by
\be \label{enewentr}
S_{BH} / N\, ,
\ee
where $S_{BH}$ is the entropy of a black hole carrying the same
set of charges as the dyon and $N$ is the order of $g$.

Given this result for the index it is natural to ask if this can
be explained from the macroscopic viewpoint. We show that it is
indeed possible to provide an explanation using the quantum entropy
function formalism\cite{0809.3304,0903.1477}, 
-- a proposal for calculating systematic quantum
corrections to the black hole entropy as a path integral of string
theory over the near horizon geometry of the black hole. We find that
when we follow the same prescription to compute the 
twisted index $B^g_6$, the path integral must be carried out over
field configurations satisfying $g$-twisted boundary condition along the
boundary circle of the $AdS_2$ factor of the near horizon geometry.
Since the circle is contractible in the interior of $AdS_2$, this is
not an allowed boundary condition on the fields in the attractor
geometry. As  a result
the saddle point corresponding to the attractor geometry does
not contribute to the path integral. 
However a $\ZZZ_N$ orbifold of the attractor geometry,
which has the same asymptotics as the attractor geometry, 
does contribute and gives a contribution to the path integral whose
semiclassical value is $\exp(S_{BH}/N)$. This provides a natural
explanation for the microscopic result \refb{enewentr}.

The rest of the paper is organised as follows. In \S\ref{ss2}
we consider the simple example of type II string theory on
$K3\times T^2$ and compute the index $B^g_6$ 
for a $\ZZZ_2$ transformation
$g$ that acts geometrically on $K3$ and
commutes with all the symmetries. The result is expressed
as a triple Fourier integral of a `partition function'. In 
\S\ref{ss3} we study various properties of this partition function
by relating it to a threshold 
integral\cite{dixon,9512046,9607029,0602254,0603066}.
In particular we show that it transforms as a modular form under
a subgroup of $Sp(2,\ZZZ)$. We also determine the location
of its zeroes and poles. In \S\ref{eind} we use these
properties to derive properties of the index. In particular
we prove the S-duality invariance of the index and show that
the jump in this index across a wall of marginal stability is
controlled by the residues at the poles of the partition
function. We also determine the behaviour of the index for
large charges, and find that the `entropy', defined
as the logarithm of the index, is half of the entropy of a black
hole carrying the same charges. In \S\ref{ss5} we generalize these
results to type IIB string theory on 
$\MM\times T^2$ where $\MM=K3$ or $T^4$, 
with $g$ chosen as a $\ZZZ_N$
transformation in $\MM$ which preserves 16 of the
supersymmetries of the theory. In this case the `entropy' associated
with the index grows as $1/N$ times the entropy of the black hole.
In \S\ref{ss6} we provide a
macroscopic explanation of this phenomenon using quantum
entropy function formalism. 
We end in \S\ref{s7} by discussing some other possible
applications of this twisted index.

\sectiono{Computation of a $\ZZZ_2$ Twisted
Index in type II String Theory 
on $K3\times T^2$
} \label{ss2}

In this section 
we shall consider the spectrum of 1/8 BPS dyons
in type IIB string theory on $K3\times T^2$, identify $g$ as a
specific geometric
$\ZZZ_2$ symmetry of $K3$ 
used in \cite{0510147,0605210,0708.1270} that preserves 
the covariantly constant spinors of $K3$ and leaves invariant
$14$ of the $22$ 2-cycles of $K3$,
and calculate the index $B^g_6$ for these dyons.
This of course forces the moduli of $K3$ to lie in a subspace
of the full moduli space admitting this symmetry.
We shall denote the $a$ and $b$ cycles of $T^2$ by $S^1$ and
$\wt S^1$, and focus on a specific class of dyons consisting of
one   D5-brane wrapped along $K3\times S^1$,
$(Q_1+1)$  D1-branes 
wrapped along $S^1$ and one Kaluza-Klein (KK)
monopole associated with the circle $\wt S^1$,  carrying
$-n$ units of momentum along
$S^1$ and $J$ units of momentum along $\wt S^1$.\footnote{The
analysis can be generalized to the case of multiple $(Q_5$) D5-branes
following \cite{0605210}. In any case the final result depends only
on the combination $Q_5(Q_1-Q_5)$.} 
A duality map
involving an S-duality of the ten dimensional
type IIB string theory, followed by a T-duality along $\wt S^1$
and finally a string string duality transformation that
relates type IIA string theory on $K3$ to heterotic string
theory of $T^4$,
brings this state to a specific state in heterotic string theory.
Under this duality map the charges $Q_1$,  $J$
and the single D5-brane wrapping along $K3\times S^1$ 
become components
of the magnetic charge $P$,
the charges $n$ and the single KK-monopole charge
associated with $\wt S^1$ become components of the electric
charge $Q$, and the transformation $g$ 
gets mapped to a specific symmetry of the heterotic string theory that
exchanges the two $E_8$ factors of the gauge 
group. 
Thus it
acts only on the left-moving modes of the world-sheet and
preserves all the 
supersymmetries.
The duality group of the theory is
$SO(6,22;\ZZZ)_T \times SL(2,\ZZZ)_S$ 
where the subscripts $T$ and $S$
denote that in this heterotic frame they 
appear as T- and S-duality
symmetries respectively. If we denote by 
$Q^2$, $P^2$ and $Q\cdot P$
the $SO(6,22)$ invariant bilinears in the charges 
then for this particular charge vector
$Q^2$, $P^2$ and $Q\cdot P$ are given by the 
relations\cite{0605210,0708.1270}
\be \label{ere1}
Q^2 = 2n, \quad P^2 = 2 Q_1, \quad Q\cdot P = J\, .
\ee
We define the partition function $Z(\rho,\sigma, v)$ via the relation:
\be \label{ere2}
Z(\rho, \sigma, v) = \sum_{n, Q_1, J} B^g_6(n, Q_1, J)\, (-1)^{J}
e^{2\pi i Q_1 \sigma
+ 2\pi i n \rho + 2\pi i J v}\, ,
\ee
where $B^g_6(n, Q_1, J)$ is the contribution to 
$B^g_6$ from states
carrying quantum numbers $(n, Q_1, J)$. Inverting this relation we
get
\be \label{ere3}
-B^g_6(n, Q_1, J) = (-1)^{J+1}\,
\int_0^1 d\rho \int_0^1 d\sigma \int_0^1 dv\, 
e^{-2\pi i Q_1 \sigma
- 2\pi i n \rho - 2\pi i J v}\, Z(\rho, \sigma, v)\, .
\ee

The computation of the $Z(\rho, \sigma, v)$ 
proceeds as in \cite{0605210}
where the $B_6$ index of dyons was calculated in 
type IIB string theory on $K3\times T^2$, modded out by a $\ZZZ_2$
transformation that involved the same symmetry $g$ accompanied by
half unit of translation along $S^1$. 
We shall follow the notations of
\cite{0708.1270} where
these results were reviewed.
The main 
difference between our analysis
and the one given in \cite{0605210} will be that 1) here we do not
remove any mode since we are not considering an orbifold, 
and 2) instead
of correlating the momentum along $S^1$ with the $g$ quantum
number as in
\cite{0605210}, here we use the $g$ quantum number of a mode 
as the weight of its contribution to the index. However as far as
the counting of the modes and their bahaviour under $g$ transformation
are concerned, we can directly use the results 
of \cite{0605210}.
As in the case of \cite{0605210} we shall 
express $Z$ as a product
of several independent pieces:
\begin{enumerate}
\item Partition function for the excitation modes of the KK 
monopole.
\item Partition function for the D1-D5 center of mass motion in the
KK monopole background. This can be further divided into the
contribution from the zero modes and the contribution from the
non-zero modes.
\item Partition function associated with the motion of the D1-branes
relative to the D5-brane.
\end{enumerate}
The world volume degrees of freedom of the three component
systems listed above, and the quantum numbers carried by them,
including their transformation properties under $g$, can be
found in \cite{0605210,0708.1270}.
As in \cite{0605210,0708.1270},
we shall work in the convention where the four $g$-invariant
unbroken supersymmetries
of the system act on the right-moving modes, \i.e.\ modes carrying
positive momentum along $S^1$. In this 
convention the requirement of
unbroken supersymmetry forces all the 
right-moving modes into their ground states.

In the process of computing the index we must make sure that
all the $g$-invariant
fermion zero modes are absorbed by the factors of $2h$ inserted
into the trace. Since in the present example $g$ commutes with
all the supersymmetries of the theory, all the fermion zero
modes associated with the broken supersymmetries will be
$g$ even.
In particular the KK monopole world-volume
breaks 8 of the 16 supersymmetries, producing 8 fermion zero
modes.
These
eight $g$-even fermion zero modes 
are soaked up by 4 factors of $2h$ in $B^g_6$. The D1-D5 system
in the KK monopole background 
breaks four more supersymmetries.
This exhausts all the fermion zero modes associated with broken
supersymmetry generators as long as we consider only BPS excitations,
\i.e.\ excitations carrying negative momentum along $S^1$.

We begin by computing
the contribution to the partition function
from the KK monopole. 
It follows from the results reviewed in \cite{0708.1270}
that the world-volume of the KK monopole has 
16 left-moving bosonic oscillators even under $g$ and 8 left-moving
bosonic oscillators odd under $g$.
None of these excitations carry any D1-brane 
charge or momentum along
$\wt S^1$. Furthermore the ground state of the Kaluza-Klein
monopole carries $-1$ unit of left-moving momentum, \i.e.\
one unit of momenum along $S^1$. 
Thus
the net contribution to the partition function from these modes is
given by
\be \label{ere5}
Z_{KK} = e^{-2\pi i\rho}
\, \prod_{n=1}^\infty \left( 1 - e^{2\pi i n\rho}\right)^{-16}\, 
\prod_{n=1}^\infty \left( 1 + e^{2\pi i n\rho}\right)^{-8}\, .
\ee
There are also eight right-moving fermion zero modes which
are absorbed by four factors of $2h$ inserted into the trace.

Next we turn to the contribution from the D1-D5 center of mass
motion in the background of the KK monopole. 
Again the various modes and their quantum numbers can be
read out from the results reviewed in \cite{0708.1270}.
In particular all of these modes are $g$-invariant.
There are four fermion zero modes associated with broken
supersymmetry; they are absorbed by two factors of $2h$
inserted into the helicity trace.
The dynamics of the rest of the
zero modes is described by an
interacting supersymmetric quantum mechanics
with Taub-NUT target space, and using the results of
\cite{pope,9912082} one finds that the
corresponding contribution to the partition function is given by
$-e^{-2\pi i v} / (1-e^{-2\pi i v})^2$\cite{0708.1270}. 
On the other hand the
non-zero mode oscillators consist of four 
$g$-invariant left-moving 
bosonic modes
carrying $\pm 1$ units of momentum along $\wt S^1$, and 
four $g$-invariant left-moving
fermionic modes carrying no momentum 
along $\wt S^1$.\footnote{At the center of the KK monople
the quantum number labeling $\wt S^1$ momentum can be identified
as the $U(1)_L$ generator of the $SU(2)_R\times SU(2)_L$
rotation group in the tangent space of the KK monopole.}
Thus the net contribution to the partition
function from the zero modes and the non-zero mode oscillators
associated with the D1-D5-brane center of mass motion is given by
\be \label{edcm}
Z_{cm}=  -e^{-2\pi i  v} \,  (1 - e^{-2\pi i  v})^{-2}\,
 \prod_{n=1}^\infty \left\{
(1 - e^{2\pi i n  \rho})^4 \, ( 1 - e^{2\pi i n  \rho + 2\pi i
 v})^{-2} \,  ( 1 - e^{2\pi i n \rho - 2\pi i
 v})^{-2}\right\}\, .
\ee

Finally we turn to the contribution from the motion of the D1-branes
relative to the D5-brane. 
For this we first need to introduce
some auxiliary quantities. The low energy dynamics of a single
D1-brane inside the D5-brane, wound once along $S^1$, is
described by a superconformal field theory with target space
$K3$. We define\cite{0708.1270}
\be\label{esi4aint}
F^{(r,s)}(\tau,z) \equiv {1\over 2} Tr_{RR; g^r} \left(  g^s
(-1)^{J_L+ J_R}
e^{2\pi i \tau L_0}
e^{-2\pi i \bar\tau \bar L_0} 
e^{2\pi i J_L z}\right), \qquad r,s=0,1\, ,
 \ee
where
$Tr$ denotes trace over all the $g^r$ twisted
Ramond-Ramond (RR) sector 
states in this CFT,  
and $J_L/2$ and $J_R/2$ denote the generators of the
$U(1)_L\times U(1)_R$ subgroup of the $SU(2)_L\times SU(2)_R$
R-symmetry group of this conformal field theory.
In
our convention the current associated with $J_L$ is holomorphic
and the one associated with $J_R$ is anti-holomorphic.
Explicit computation gives\cite{0602254}
\ben\label{valind}
&& F^{(0,0)}(\tau, z) = 4\left[ {\vartheta_2(\tau,z)^2
\over \vartheta_2(\tau,0)^2} +
{\vartheta_3(\tau,z)^2\over \vartheta_3(\tau,0)^2}
+ {\vartheta_4(\tau,z)^2\over \vartheta_4(\tau,0)^2}\right]\, ,
\nonumber \\
&& F^{(0,1)}(\tau, z) =  4 \, {\vartheta_2(\tau,z)^2\over
\vartheta_2(\tau,0)^2} \, , \quad 
F^{(1,0)}(\tau, z) = 4 {\vartheta_4(\tau,z)^2\over \vartheta_4(\tau,0)^2}
\, ,
\quad 
F^{(1,1)}(\tau, z) =  4\, {\vartheta_3(\tau,z)^2\over
\vartheta_3(\tau,0)^2} \, ,\nonumber \\
\een
where $\vt_i$ are the Jacobi theta functions.
\refb{valind} can be rewritten as
\be\label{ee3}
F^{(r,s)}(\tau, z) = h_0^{(r,s)}(\tau) 
 \, \vartheta_3(2\tau, 2z)
+ h^{(r,s)}_1(\tau) 
\, \vartheta_2(2\tau, 2z) 
\ee
where
\ben\label{defhlrs}
&& h^{(0,0)}_0(\tau) = 
8\, {\vartheta_3(2\tau,0)^3 \over \vartheta_3(\tau,0)^2
\vartheta_4(\tau,0)^2} + 2\, {1\over \vartheta_3(2\tau,0)}
\nonumber \\
&& h^{(0,0)}_1(\tau) = -8\, {\vartheta_2(2\tau,0)^3 
\over  \vartheta_3(\tau,0)^2
\vartheta_4(\tau,0)^2} + 2\, {1\over \vartheta_2(2\tau,0)}
\nonumber \\
&& h_0^{(0,1)}(\tau) = 2\, {1\over \vartheta_3(2\tau,0)}, \qquad
h_1^{(0,1)}(\tau) = 2\, {1\over \vartheta_2(2\tau,0)},
\nonumber \\
&& h_0^{(1,0)}(\tau) = 4 {\vartheta_3(2\tau, 0)\over
\vartheta_4(\tau,0)^2}, \qquad
h_1^{(1,0)}(\tau) =  -4 {\vartheta_2(2\tau, 0)\over
\vartheta_4(\tau,0)^2},
\nonumber \\
&& h_0^{(1,1)}(\tau) = 4 {\vartheta_3(2\tau, 0)\over
\vartheta_3(\tau,0)^2}, \qquad
h_1^{(1,1)}(\tau) =  4 {\vartheta_2(2\tau, 0)\over
\vartheta_3(\tau,0)^2}\, .
\een
We now define the coefficients $c^{(r,s)}_b(u)$ through the expansions
\be \label{defcrs}
h_b^{(r,s)} (\tau) = \sum _{n\in {1\over 2}\zzz -{1\over 4}
b^2} c^{(r,s)}_b(4n) q^n, \qquad b=0,1\, .
\ee
Substituting \refb{defcrs}
into \refb{ee3} and using the Fourier expansions of $\vt_3(2\tau,2z)$,
$\vt_2(2\tau,2z)$ we get
\be\label{e1.5copy}
F^{(r,s)}(\tau,z) =\sum_{b=0}^1
\sum_{j\in2\zzz+b, n\in {1\over 2}\zzz} c_b^{(r,s)}(4n -j^2)
e^{2\pi i n\tau + 2\pi i jz}\, .
\ee

Consider now the motion of a single D1-brane, wound $w$ times
along $S^1$,  inside a D5-brane. 
The dynamics of this system is described by a superconformal
field theory with target space $K3$, but since the D1-brane has
length $w$ times the period of $S^1$, one unit of left-moving
momentum along
$S^1$ will appear as $w$ units of left-moving
momentum ($L_0$) on the D1-brane.
Also in the background of the Kaluza-Klein monopole the
quantum numbers $(J_L, J_R)$ can be identified respectively with
the $\wt S^1$ momentum $J$ and the fermion number $F$ of the
four dimensional theory\cite{0503217,0505094,0508174}.
We denote by $(-1)^j\, 
n(w,l, j; k)$ the
total number of bosonic minus fermionic 
states of this D1-brane, carrying $g$ quantum number $(-1)^k$,
momentum $-l$ along $S^1$ and momentum $j$ along $\wt S^1$.
Then it follows from \refb{esi4aint}, \refb{e1.5copy}
 that\cite{0605210,0708.1270}\footnote{The main 
difference between our results
and those reviewed 
in \cite{0708.1270} is that in \cite{0708.1270} the
$g$ transformation law was correlated with the momentum along
$S^1$. In our case they are independent data.}
\be \label{enwj}
n(w,l, j; k) = \sum_{s=0}^{1}\,  (-1)^{sk} 
\, c_b^{(0,s)}(4lw-j^2)\, , \quad l,w,j\in\ZZZ\, ,\quad b=j \, \hbox{mod}\,
2, 
 \quad l\ge 0, \quad w\ge 1\, .
\ee
Using this and the techniques of \cite{9608096}
we can compute the contribution to the partition function
from the general motion of the D1-branes inside a D5-brane, where
we have excitations involving multiple D1-branes 
carrying different values of $w$, $l$
and $j$. This is given by
\ben \label{ezd1d5}
Z_{D1D5}&=& e^{-2\pi i \sigma}\, 
\prod_{w=1}^\infty \prod_{l=0}^\infty \prod_{j\in \zzz}
\prod_{k=0}^1 \, \left( 1 - (-1)^k \, e^{2\pi i (w\sigma +
l\rho + j v)}\right)^{-n(w,l,j;k)} \nonumber \\
&=& e^{-2\pi i\sigma}\,
\prod_{b=0}^1\prod_{w=1}^\infty \prod_{l=0}^\infty 
\prod_{j\in 2\zzz+b}
\prod_{k=0}^1 \, \left( 1 - (-1)^k \, e^{2\pi i (w\sigma +
l\rho + j v)}\right)^{-\sum_{s=0}^{1}\,  (-1)^{sk} 
\, c_b^{(0,s)}(4lw-j^2)}
\, .\nonumber \\
\een
The extra factor of $e^{-2\pi i\sigma}$ reflects that the total number
of D1-branes is $Q_1+1$ and not $Q_1$. This extra shift in
$Q_1$ by 1 accounts for 
the fact that a single D5-brane wrapped on $K3$ carries $-1$
unit of D1-brane charge.
Using the results
\ben\label{ecrsv}
&& c_0^{(0,0)}(0) = 10\, , \qquad 
c_1^{(0,0)}(-1) = 1\, , \qquad c_0^{(0,1)}(0) = 2\, , 
\qquad c_1^{(0,1)}(-1) = 1\, ,  \nonumber \\
&& c_0^{(1,0)}(0)=4, \qquad c_1^{(1,0)}(-1) = 0, 
\qquad c_0^{(1,1)}(0) = 4, 
\qquad c_1^{(1,1)}(-1)=0\, ,
\een
we can express the total partition function as 
\ben \label{etotalz}
Z(\rho,\sigma,v) &=& Z_{KK} Z_{cm}  Z_{D1D5}
= 1 / \Phi(\rho, \sigma, v)\, , \nonumber \\
\Phi(\rho, \sigma, v) &=& e^{2\pi i (\rho+\sigma+v)}
\prod_{w=0}^\infty \prod_{l=0}^\infty 
 \prod_{j\in \zzz\atop j<0 \, for\, w=l=0}
\prod_{k=0}^1 \, \left( 1 - (-1)^k \, e^{2\pi i (w\sigma +
l\rho + j v)}\right)^{\sum_{s=0}^{1}\,  (-1)^{sk} 
\, c_b^{(0,s)}(4lw-j^2)}
\, ,\nonumber \\
&& \qquad \qquad \qquad \qquad \qquad \qquad \qquad \qquad
b =j \, \hbox{mod} \, 2\, . \een
The $w=0$ term in the above product is obtained from the
$Z_{KK}Z_{cm}$ term computed from
\refb{ere5}, \refb{edcm}.

\sectiono{Properties of the Dyon Partition Function} \label{ss3}

Various symmetries of the dyon partition function are best studied by
relating the function $\Phi$ defined in \refb{etotalz} to a
threshold 
integral\cite{dixon,9512046,9607029,0602254,
0605210,0607155}. Most
of the results that we shall be using can be found in appendices
C and D of \cite{0708.1270} (the function $\Phi$ was called $\wh\Phi$
in \cite{0708.1270}).
We begin 
by defining:
\be\label{edefomega}
\Omega=\pmatrix{\rho  & v \cr v  & \sigma}\, ,
\ee
and
\bea{e7n}
{1\over 2} p_R^2 &=& {1\over 4 \det Im  \Omega} |-m_1 \rho  +
m_2 + n_1 \sigma + n_2 (\sigma\rho -v^2) + j v |^2, \nonumber \\
{1\over 2} p_L^2
&=& {1\over 2}  p_R^2 + m_1 n_1 + m_2 n_2 + {1\over 4} j^2\, .
\eea
We now consider the 
integral\cite{0708.1270}:
\be\label{rwthrint}
\hI(\rho , \sigma, v)
= \sum_{b=0}^1\sum_ {r, s =0}^{1}\, \hI_{r,s,b}\, ,
\ee
where
\be\label{enn6}
\hI_{r,s,b} = \int_{\FF} \frac{d^2\tau}{\tau_2} 
\left[\sum_{\stackrel{m_1, n_1\in \zzz, m_2\in 
\zzz/2}{n_2\in 2\zzz+ r, 
j\in 2\zzz+b}}
q^{p_L^2/2} \bar q^{ p_R^2/2} (-1)^{2 m_2 s} 
h_b^{(r,s)}(\tau)-  \delta_{b,0}\, \delta_{r,0} \, c_0^{(0,s)}(0)
\right]\, ,
\qquad
q\equiv e^{2\pi i\tau}\, .
\ee
$\FF$ denotes the fundamental region of $SL(2,\ZZZ)$ in the
upper half plane. The subtraction terms 
proportional to $c_0^{(0,s)}(0)$ have been chosen so that the
integrand vanishes faster than $1/\tau_2$
in the $\tau\to i\infty$ limit, rendering the
integral finite.
Following the procedure described in \cite{0602254} 
one can show that\cite{0708.1270}\footnote{Note that although
the definition of $\Phi$ contains only the coefficients 
$c^{(0,s)}_b$, $\hI$ given in \refb{rwthrint}, \refb{enn6}
contains $c^{(r,s)}_b$ for all $(r,s)$. This is due to the fact that
the manipulations leading to \refb{enn9} makes use of the
modular property of $h^{(r,s)}_b(\tau)$ and such modular
transformations give $h_b^{(0,s)}(\tau)$ in terms of
$h_b^{(r',s')}(\tau')$ for all $(r',s')$.}
\be\label{enn9}
\hI(\rho ,\sigma, v ) = -2 \ln \left[
 ( \det\,{\rm Im}\Omega)^k \right] - 2\ln \Phi (\rho ,\sigma,v )
- 2\ln \overline{{\Phi}(\rho ,\sigma,v ) }+\hbox{constant}
\ee
where bar denotes complex conjugation,
\be\label{ekvalue}
k={1\over 2}\, \sum_{s=0}^{1} \, c_0^{(0,s)}(0) = 6\, ,
\ee
and $\Phi$ is the same function that appears in 
the expression \refb{etotalz} for
the partition function. 

Due to the relation \refb{enn9}, symmetries of $\Phi$ can
be determined from the symmetries of 
$\hI$\cite{0602254}.\footnote{Under some
transformations $\hI$ may remain invariant but $\Phi$ may pick
up a phase, but one can show that for the symmetries which will be
relevant for our discussion this does not happen.}
Consider in particular $O(3,2;\ZZZ)$ transformation on the variables
$(\rho,\sigma, v)$ and 
$(m_1, n_1, m_2, n_2, j)$ defined as follows:
\be\label{eso32}
\pmatrix{m_1'\cr m_2'\cr n_1'\cr n_2'\cr j'}
= S \pmatrix{m_1\cr m_2\cr n_1\cr n_2\cr j}\, ,
\qquad \pmatrix{\sigma'\cr \rho' \sigma'-v^{\prime2}\cr -\rho'
\cr 1\cr 2v'}=
\lambda \, S\, \pmatrix{\sigma \cr \rho \sigma-v^2\cr -
\rho\cr 1\cr 2v}\, 
\ee
where $S$ is a $5\times 5$ matrix with integer entries, satisfying
\be\label{eso33}
S^T L S=L, \qquad L = \pmatrix{0 & I_2 & 0\cr I_2 & 0 & 0\cr
0 & 0 & {1\over 2}}\, ,
\ee
and $\lambda$ is a number to be adjusted so that the fourth element of
the vector on the left hand side of (\ref{eso32}) is 1. $I_n$ denotes
$n\times n$ identity matrix.
One can easily check that $p_R^2$ and $p_L^2$ are invariant
under these transformations. Thus as long as the transformation
\refb{eso32} preserves the restriction on 
$(m_1, n_1, m_2, n_2, j)$ in the sum in \refb{enn6}, 
and preserves $m_2$ mod 1, this transformation
is a symmetry of $\hI$, and hence of 
$(\det Im\Omega)^k\Phi\bar\Phi$. 
{}From the known modular transformation properties of 
$\det Im\Omega$
it then follows that $\Phi$ transforms as a modular form of weight
$k$ under these transformations.
Since $O(3,2;\ZZZ)$ is
isomorphic to the modular group $Sp(2;\ZZZ)$ of genus two
Riemann surfaces we see that $\Phi$ is a
modular form of a subgroup of $Sp(2,\ZZZ)$. The generators of this
subgroup have been given explicitly 
in \cite{0510147,0708.1270}.

A special subgroup of the symmetry group is generated by the following
transformations:
\ben \label{etrs1}
(m_1, n_1, m_2, n_2, j) &\to& (-n_1, -m_1, m_2, n_2, -j),  \nonumber \\
\hbox{and}
\quad 
(m_1, n_1, m_2, n_2, j) &\to& (m_1 - n_1 - j, n_1, m_2, n_2, j+ 2 n_1)\, .
\een
Via \refb{eso32} these induce the symmetries
\be \label{etrs2}
\Phi(\rho,\sigma, v) =\Phi(\sigma, \rho, -v), \qquad \hbox{and}
\qquad
\Phi(\rho,\sigma, v) =\Phi(\rho,\sigma+\rho-2v, v-\rho) \, .
\ee
As we shall see in \S\ref{ss3.5}, these generate
the S-duality symmetry of the partition function $\Phi^{-1}$.

Eq.\refb{enn9} also allows us to determine the zeroes and poles
of $\Phi$, since they correspond
to logarithmically divergent contribution
to the threshold integral $\wh \II$. A detailed analysis can be found
in \cite{0605210,0708.1270}; here we summarize the main results. 
Analyzing \refb{enn6} one can show that $\hI$ can get 
logarithmically divergent
contributions when $p_R^2$ vanishes and $p_L^2={1\over 4}$ 
 for one of the terms
in the sum, \i.e.\ near the point
\be \label{elog1}
 -m_1 \rho  +
m_2 + n_1 \sigma + n_2 (\sigma\rho -v^2) + j v =0\, ,
\qquad m_1 n_1 + m_2 n_2 +{j^2\over 4} = {1\over 4}\, ,
\ee
where $(m_1, n_1, m_2, n_2, j)$ takes one of the values which
appear in the sum in \refb{enn6}.
Near such a point $\Phi(\rho, \sigma, v)$ behaves as\cite{0708.1270}
\ben \label{ebeh1}
\Phi(\rho, \sigma, v) &\sim& ( -m_1 \rho  +
m_2 + n_1 \sigma + n_2 (\sigma\rho -v^2) + j v)^{\sum_{s=0}^1
(-1)^{2\, m_2 s}
c_1^{(r,s)}(-1)}, \nonumber \\
&& \qquad r = n_2 \quad \hbox{mod} \quad 2, \quad
m_1, n_1, n_2 \in \ZZZ, \quad m_2\in \ZZZ/2\, .
\een
Now from \refb{ecrsv} we see that $c_1^{(1,s)}(-1)=0$ for all $s$.
Thus there are no poles and zeroes of $\Phi$ for odd $n_2$.
On the other hand for $n_2$ even we have $r=0$ in \refb{ebeh1},
and from \refb{ecrsv} we get
\be \label{eexpo}
\sum_{s=0}^1(-1)^{2 \, m_2 \,  s}
c_1^{(0,s)}(-1)= 1+(-1)^{2 \, m_2}\, . 
\ee
Thus $\Phi$ has a
second order zero for even values of $2 m_2$ in \refb{ebeh1}
and no poles.

\sectiono{Properties of the Index} \label{eind}

In this section we shall make use of the various properties of the dyon
partition function derived in \S\ref{ss3} to derive properties of
the index $B^g_6$. In particular we shall study the properties of
$B^g_6$ under S-duality, study the jump in $B^g_6$ under wall
crossing and study the asymptotic growth of $B^g_6$ for large
charges. Our starting point will be \refb{ere3}, but using
\refb{ere1} we shall express it as:
\be \label{ere3rpt}
-B^g_6(Q,P) = (-1)^{Q\cdot P+1}\,
\int_C  d\rho \, d\sigma \, dv\, 
e^{-2\pi i \left( \sigma {P^2\over 2} + \rho{Q^2\over 2} + v Q\cdot P
\right)}
\, {1\over \Phi(\rho, \sigma, v)}\, .
\ee
The integral has been written as an integral over 
a `contour' $C$ which, according to \refb{ere3}, lies
along the
real $(\rho,\sigma,v)$ axes. 
Naively in \refb{ere3} we can fix $Im(\rho)$,  $Im(\sigma)$
and $Im(v)$ to any values we like; however in order that
\refb{ere2} converges we need to choose the imaginary parts
of $(\rho,\sigma,v)$ to lie in certain domains in $\RRR^3$.
As is well
understood by now, the choice of the imaginary parts of
$(\rho,\sigma, v)$ depend on the point in the moduli space
of the theory we are 
at\cite{0702141,0702150,0705.3874,0706.2363}
since the spectrum -- and hence the convergence property
of \refb{ere2} -- changes discontinuously as the moduli cross the
walls of marginal stability. Nevertheless one finds that the
function $Z(\rho,\sigma,v)$ defined through \refb{ere2}
is the same -- or more precisely the analytic continuation
of the same meromorphic function -- irrespective of where in the
moduli space we compute it. The logic leading to this conclusion
will be reviewed at the end of \S\ref{ss3.5}. 
This is equivalent to the statement 
that $B^g_6$ in different domains in the moduli space
are given by \refb{ere3rpt} with the same $\Phi$, but the choice of
the contour $C$ is different in different domains in the moduli
space, dictated by the convergence of the sum in \refb{ere2}.
A convenient prescription for the choice of contour  as a function
of the moduli is\cite{0706.2363}
\ben \label{echoiceint}
Im(\sigma) &=& \Lambda \, \left({|\tau|^2\over \tau_2} +
{Q_R^2 \over \sqrt{Q_R^2 P_R^2 - (Q_R\cdot P_R)^2}}\right)\, ,
\qquad 0 \le Re(\sigma)\le 1\, ,
\nonumber \\
Im(\rho) &=& \Lambda \, \left({1\over \tau_2} +
{P_R^2 \over \sqrt{Q_R^2 P_R^2 - (Q_R\cdot P_R)^2}}\right)\, ,
\qquad 0 \le Re(\rho)\le 1\, ,\nonumber \\
Im(v) &=& -\Lambda \, \left({\tau_1\over \tau_2} +
{Q_R\cdot P_R \over \sqrt{Q_R^2 P_R^2 - (Q_R\cdot P_R)^2}}
\right)\, ,\qquad 0 \le Re(v)\le 1\, ,
\een
where $\Lambda$ is a large positive number, 
\be \label{edefcrint}
Q_R^2 = Q^2 + Q^T MQ, \quad P_R^2=P^2 + P^T MP, \quad
Q_R\cdot P_R = Q\cdot P+ Q^T M P\, ,
\ee
$\tau\equiv \tau_1+i\tau_2$ denotes the asymptotic value
of the axion-dilaton moduli which
belong to the gravity multiplet and $M$ is the 
asymptotic value of the symmetric $O(6,22)$ matrix
valued moduli field of the matter multiplet, -- in the heterotic
description these represent the moduli of $T^6$ and Wilson
lines along $T^6$.
The choice \refb{echoiceint} 
of course is not unique since we can deform the contour
without changing the result for the index as long as we do not
cross a pole of the partition function. 
However \refb{echoiceint} gives a useful bookkeeping device
for associating domains in the moduli space to domains in the
$(Im(\rho), Im(\sigma), Im(v))$ space. Under small deformations
of the moduli $(\tau,M)$, \refb{echoiceint} induces small deformations
of the contour $C$. 
While generically the value of the integral
does not change under such small deformations, the
result does change if the contour \refb{echoiceint}
crosses a zero of $\Phi$.
Physically these jumps are associated with the jumps in the index
across walls of marginal stability\cite{0702141,0702150}.

Using T-duality symmetry of the theory one can argue that any other
charge vector that can be related to the D1-D5-KK monopole system
considered here via a $g$-invariant T-duality transformation will have
$B^g_6$ given by \refb{ere3rpt}. The choice of contour given in
\refb{echoiceint} is manifestly invariant 
under T-duality transformation.

\subsection{S-duality Invariance} \label{ss3.5}

S-duality transformations act on the charges and moduli as
\be \label{esdtrs}
\pmatrix{Q\cr P} \to \pmatrix{Q'\cr P'}
=\pmatrix{a & b\cr c & d}
\pmatrix{Q\cr P}\, , \quad 
\tau\to \tau'={a\tau +b\over c\tau + d}, \quad M\to M\, ,
\quad \pmatrix{a & b\cr c & d}\in SL(2,\ZZZ)\, .
\ee
We want to show that \refb{ere3rpt}, with the choice of
contour given in \refb{edefcrint}, is invariant under this transformation.
Let us define
\ben \label{e3.3}
     \sigma' = a^2 \sigma + b^2 \rho - 2 a
    b   v \, ,
   \nonumber \\
    \rho' = c^2 \sigma + d^2  \rho - 2
   c d  v \, ,
   \nonumber \\
     v' = - a c \sigma -  b 
   d  \rho + ( a d +  b c)  v \, .
\een
One can easily verify that
\be\label{e3.4}
e^{-\pi i ( \rho \, Q^2
+ \sigma \,P^2  +2 v \, Q\cdot P)} = 
e^{-\pi i ( \rho' Q^{\prime 2}
+ \sigma' P^{\prime 2} +2 v' 
Q'\cdot P')}   \, ,
\ee
and
\be\label{e3.5}
d \rho \, 
d \sigma \,
d v = d \rho' \, 
d \sigma' \,
d v' \, .
\ee
Furthermore,
using the results
$Q^2,P^2\in 2\ZZZ$ and $Q\cdot P\in\ZZZ$ 
one finds that $(-1)^{Q\cdot P}
=(-1)^{Q'\cdot P'}$.
Using these relations
we can rewrite \refb{ere3rpt} as
\be \label{ebgtrs}
B^g_6(Q,P;\tau,M) = (-1)^{Q'\cdot P'+1}\,
\int_C\,  d\rho'  d\sigma' \, dv'\, 
e^{-2\pi i \left( \sigma' {P^{\prime 2}\over 2} 
+ \rho'{Q^{\prime 2}\over 2} + v' Q'\cdot P'
\right)}
\, {1\over \Phi(\rho, \sigma, v)}\, .
\ee
Note that we have now explicitly indicated the dependence of $B^g_6$
on the moduli $\tau$, $M$ via the choice of the contour 
\refb{echoiceint}.
One can now further observe that
\begin{itemize}
\item The choice of the contour $C$ given in \refb{echoiceint} is
S-duality covariant under simultaneous S-duality transformation 
on $(\tau, Q, P)$ given in \refb{esdtrs} and of $(\rho,\sigma, v)$
given in \refb{e3.4}. Thus we can replace $C$ by $C'$ in 
\refb{ebgtrs} with the understanding that $C'$ corresponds to the
contour where all the variables are replaced by primed variables
in \refb{echoiceint}.

\item We have the relation
\be \label{ephtrs}
\Phi(\rho, \sigma, v) = \Phi(\rho', \sigma', v')\, .
\ee
This follows from the fact that \refb{etrs2} corresponds to
invariance of $\Phi$ under \refb{e3.3} for
\be \label{ephsym}
\pmatrix{a & b\cr c & d} = \pmatrix{0 & 1\cr -1 & 0}, \quad
\hbox{and} \quad \pmatrix{a & b\cr c & d} = \pmatrix{1 & 1\cr 0 & 1}
\, ,
\ee
and that the two matrices in \refb{ephsym} generate the whole
S-duality group.
\end{itemize}
Using these two results we can rewrite \refb{ebgtrs} as
\ben \label{ebgtrsnew}
-B^g_6(Q,P;\tau,M) &=& (-1)^{Q'\cdot P'+1}\,
\int_{C'}\,  d\rho'  d\sigma' \, dv'\, 
e^{-2\pi i \left( \sigma' {P^{\prime 2}\over 2} 
+ \rho'{Q^{\prime 2}\over 2} + v' Q'\cdot P'
\right)}
\, {1\over \Phi(\rho', \sigma', v')}\nonumber \\
&=& -B^g_6(Q',P';\tau',M)\, .
\een
This finishes the proof of S-duality invariance of
$B^g_6$.

In practice we derive the expression \refb{ere3rpt}, 
\refb{echoiceint} for the
index only in certain domains in the moduli space where the
type IIB string theory is weakly coupled\cite{0605210,0708.1270}
and then extend the
result to other domains by {\it requiring S-duality invariance
of the spectrum\cite{0702141,0702150}.} 
Thus our analysis in this section should really
be regarded as a proof not of S-duality invariance but of
\refb{ere3rpt}, \refb{echoiceint}, 
-- \i.e.\ of the statement
that the same function $\Phi$ can be used
to capture the index in different domains in the moduli space
just by changing the contour according to \refb{echoiceint}.

\subsection{Wall Crossing} \label{ss4}

If we keep the charges fixed and vary the asymptotic moduli then the
integration contour $C$ varies. When it hits a pole of the integrand
there is a jump in $B^g_6$ given by the residue of the integrand
at the pole. 
The physical interpretation of this jump is that at these points
in the moduli space we have a wall of marginal stability and the
jump in the degeneracy is due to the jump in the index across the
wall of marginal stability\cite{0702141,0702150}.
One can show that\cite{0702141} the poles which are encountered
in this process are of the form
\be \label{epolestr}
\sigma \gamma - \rho \beta + v (\alpha-\delta) = 0\, ,
\qquad \alpha\delta=\beta\gamma, \qquad \alpha+\delta=1,
\qquad \alpha,\beta,\gamma, \delta \in \ZZZ\, .
\ee
The corresponding wall of marginal stability
is associated with the decay
\be \label{edecay}
(Q,P) \to (\alpha Q +\beta P, \gamma Q +\delta P)
+
(\delta Q -\beta P, -\gamma Q + \alpha P)\, .
\ee
In fact via S-duality transformation 
all of these decays can be related
to the decay\cite{0702141}
\be \label{esimpled}
(Q,P)\to (Q, 0) + (0, P)\, ,
\ee
and the corresponding pole of the partition function is at
\be \label{esimpole}
v = 0\, .
\ee
Using \refb{etotalz} and the identity
\be \label{eidentity}
\sum_{b=0}^1\sum_{j\in 2\zzz+b}
 c_b^{(r,s)}(4n - j^2) = \delta_{n,0}\, 
\{c_0^{(r,s)}(0) + 2\, c_1^{(r,s)}(-1)\}\, ,
\ee
we see that for small $v$
\be \label{evzero}
\Phi(\rho,\sigma,v) 
= -4\, \pi^2 \, v^2\, g(\rho) \, g(\sigma) + \OO(v^4)\, ,
\ee
\be \label{edefg}
g(\rho) \equiv  e^{2\pi i\rho}\, \prod_{n=1}^\infty 
\left( 1 - e^{2\pi i n\rho}\right)^{16} \, 
\left( 1 + e^{2\pi i n\rho}\right)^{8}\, .
\ee
Thus the jump in the index, given by the residue at the pole
of the integrand in \refb{ere3rpt} at $v=0$, is given 
by\footnote{The sign of $\Delta B^g_6$ of course depends on
in which direction the contour crosses the pole, which in turn
is determined by the direction in which the moduli cross the
wall of marginal stability.}
\be \label{ejump}
\Delta B^g_6 = (-1)^{Q\cdot P+1} (Q\cdot P) \, f(Q) \, f(P)\,,
\ee
where 
\be \label{edeffq}
f(Q) = \int_0^1 \, d\rho\, 
 e^{-i\pi  \rho Q^2} \, {1\over g(\rho)}\, .
\ee
It is easy to see that $1/g(\rho)$ is precisely the partition function
that computes the $B^g_4$ index of the dyons 
carrying charges
$(Q,0)$ (\i.e.\ the KK monopole carrying momentum along $S^1$)
or $(0,P)$ (\i.e.\ the D1-D5 system carrying momentum along $\wt S^1$)
and preserving half of the $g$-invariant supersymmetries.
For example $1/g(\rho)$ is precisely the partition function of the
Kaluza-Klein monopole given in \refb{ere5} after removing the
contribution from the fermion zero modes. Thus \refb{ejump} can be
rewritten as
\be \label{enewjump}
\Delta B^g_6 = (-1)^{Q\cdot P+1} \, (Q\cdot P) \, 
B^g_4((Q,0))\, B^g_4((0,P))\, ,
\ee
in agreement with the 
wall crossing formula for the index 
$B_6$\cite{0803.3857}.

\subsection{Asymptotic Growth} \label{ss5a}

We shall now study the asymptotic growth of the index $B^g_6$ for
large charges. 
As was shown in \cite{9607026,0412287,0510147,
0605210} and reviewed in \cite{0708.1270}, when 
$Q^2$, $P^2$ and $Q\cdot P$ are
all large and of the same order the asymptotic growth of the
index, given by \refb{ere3}, is controlled by the pole of the
partition function, \i.e.\ zeroes of $\Phi$ at
\be \label{epole}
-m_1 \rho  +
m_2 + n_1 \sigma + n_2 (\sigma\rho -v^2) + j v =0\, ,
\qquad m_1 n_1 + m_2 n_2 +{j^2\over 4} = {1\over 4}\, ,
\ee
for $|n_2|>0$. Furthermore the contribution from a pole of this type is
of order
\be \label{eorder}
\exp\left[\pi\sqrt{Q^2 P^2 - (Q\cdot P)^2} / |n_2| + \cdots\right]\, ,
\ee
where $\cdots$ denotes terms which are of order unity or suppressed
by powers of the charges. Thus the leading contribution to the entropy
comes from the poles with lowest possible non-zero
value of $|n_2|$.
Since from the analysis below \refb{ebeh1} it follows that 
$\Phi$ has no zeroes for odd $n_2$ we see that the leading
contribution to $B^g_6$ comes from the pole(s) with 
$|n_2|=2$.
Thus for large charges we have
\be \label{eentropy}
\ln \left|B^g_6\right|
 = {\pi\over 2}\sqrt{Q^2 P^2 - (Q\cdot P)^2} + \cdots\, .
\ee
Since the entropy of a BPS black hole carrying these
charges is given by
$\pi\sqrt{Q^2 P^2 - (Q\cdot P)^2}$\cite{9507090,9512031},
we see that $\ln \left|B^g_6\right|$ 
is half of the entropy of a black hole carrying the
same charges (and also half of the `entropy' computed from the usual
helicity trace index $B_{6}$). 
We shall provide a macroscopic
explanation of this phenomenon in \S\ref{ss6}. 

\sectiono{Generalization to $\ZZZ_N$ Twisted
Index} \label{ss5}

The above results can be easily generalized to the case where the
theory under consideration is type IIB string theory on $\MM\times T^2$
where $\MM$ can be either $T^4$ or $K3$,
and  the $\ZZZ_2$ symmetry generator $g$ is replaced by a
$\ZZZ_N$ symmetry generator $g$ that has a geometric action
on $K3$ or $T^4$, and commutes with an $\NN=4$ subalgebra
of the full supersymmetry algebra. This implies that
for $\MM=T^4$ $g$ commutes with half of the
32 supersymmetries, while for $\MM=K3$ $g$ 
must commute with all the 16 
supersymmetries.\footnote{Following the same set of duality
transformations as in the $\ZZZ_2$ example described earlier,
one can map this theory to type II or heterotic string theory
on $T^6$, with the $\ZZZ_N$ acting only on the left-moving
fields on the world-sheet.}
The $\ZZZ_N$ transformations we shall be using
can be found in \cite{0609109} and references therein. However unlike
in \cite{0609109}, we are not computing the spectrum in a new theory
obtained by taking a $\ZZZ_N$ orbifold of the original theory.
Our interest is to compute the $g$ twisted index $B^g_6$ in the
original theory, \i.e.\ in type IIB string theory on $T^4\times T^2$
or $K3\times T^2$.

The dyon system we consider is identical to the one described in
\S\ref{ss2}, with the only difference that for $\MM=T^4$ we denote
the number of D1-branes by $Q_1$.
The method of analysis is 
also identical to that in \S\ref{ss2} and all
the technical results needed for the computation can be found in
\cite{0609109,0708.1270}. 
The only extra complication 
for $\MM=T^4$ arises from the additional degrees
of freedom associated with the Wilson line on the D5-brane along
$T^4$, but their effect can be easily computed since the
quantum numbers and the $g$ transformation laws of these
additional degrees of freedom have been given in 
\cite{0609109,0708.1270}. 
We shall describe only the final results.
First we define
\be\label{enx1}
F^{(r,s)}(\tau,z) \equiv {1\over N} Tr_{RR; g^r} \left(  g^s
(-1)^{J_L+ J_R}
e^{2\pi i \tau L_0} e^{-2\pi i \bar\tau 
\bar L_0} e^{2\pi i J_L z}\right), \qquad 0\le r,s\le N-1, 
\quad r,s\in\ZZZ\, ,
 \ee
where
$Tr$ denotes trace over all the $g^r$ twisted
RR sector 
states in the (4,4) superconformal field theory with target
space $\MM$,  
and $J_L/2$ and $J_R/2$ denote the generators of the
$U(1)_L\times U(1)_R$ subgroup of the $SU(2)_L\times SU(2)_R$
R-symmetry group of this conformal field theory.
For $\MM=T^4$ the computation of $F^{(r,s)}$ is straightforward
since we have a free conformal field theory. For $\MM=K3$ we can
calculate $F^{(r,s)}$ by working at special points in the moduli
space {\it e.g.} at the orbifold points or the Gepner points. For
prime values of $N$ explicit expression for $F^{(r,s)}$ can be
found in \cite{0602254,0607155,0708.1270}. On general grounds
$F^{(r,s)}(\tau,z)$ can be shown to have an expansion of the
form
\be\label{enewkk}
F^{(r,s)}(\tau,z) =\sum_{b=0}^1\sum_{j\in2\zzz+b, n\in \zzz/N
\atop
4n - j^2\ge -b^2} 
c^{(r,s)}_b(4n -j^2)
e^{2\pi i n\tau + 2\pi i jz}\, .
\ee
This defines the coefficients $c^{(r,s)}_b(u)$. We also define
\be\label{eqrsrevint}
Q_{r,s} = N\, 
\left( c^{(r,s)}_0(0)+ 2 \, c^{(r,s)}_1(-1)\right)\, .
\ee
Some useful relations are
\be \label{euseful}
c_1^{(0,s)}(-1) =\cases{ {2\over N} \quad \hbox{for}\quad
\MM=K3\cr
{1\over N}\left(2 - e^{2\pi i s/N} - e^{-2\pi i s/N}
\right) \quad \hbox{for}\quad
\MM=T^4
}\, .\ee
The index $B^g_6$ is then given by
\be \label{ebg6new}
-B^g_6(Q,P) 
= (-1)^{Q\cdot P+1}\,
\int_0^1 d\rho \int_0^1 d\sigma \int_0^1 dv\, 
e^{-\pi i  \sigma \, P^2
- \pi i  \rho \, Q^2 - 2\pi i  v \, 
Q\cdot P}\, {1\over \Phi(\rho, \sigma, v)}\, ,
\ee
where
\ben\label{enn9c}
\Phi(\rho,\sigma,v) &=&  C^3\, e^{2\pi i 
\wh \alpha\left(\rho+\sigma+v\right) } \nonumber \\
&& \prod_{b=0}^1\,
\prod_{r=0}^{N-1}\,  \prod_{(k',l)\in \zzz,j\in 2\zzz+b\atop
k',l\ge 0, j<0 \, {\rm for}
\, k'=l=0}
\Big\{ 1 - e^{2\pi i r / N} \, e^{ 2\pi i ( k' \sigma + l \rho + j v) 
}
\Big\}^{ \sum_{s=0}^{N-1}
e^{-2\pi i rs/N}  c^{(0,s)}_b(4k'l - j^2)
}\, ,  \nonumber \\
\een
\be\label{enn9d}
\wh \alpha= \cases{1 \quad \hbox{for} \quad \MM=K3\cr
0 \quad \hbox{for} \quad \MM=T^4}\, ,
\qquad 
C = \cases{1 \quad \hbox{for} \quad \MM=K3\cr
\left(1-e^{2\pi i/N}\right)^{-1}
\left(1-e^{-2\pi i/N}\right)^{-1}  
\quad \hbox{for} \quad \MM=T^4}\, .
\ee
The factor of $C^{3}$ in $\Phi$ comes from the quantization of
the $g$ non-invariant fermion zero modes carrying no $J$ quantum
number.

The threshold integral that can be used to derive various properties
of $\Phi$ is given by (see appendix C of \cite{0708.1270})
\be \label{eth1}
\hI(\rho , \sigma, v)
= \sum_ {r, s =0}^{N-1}\sum_{b=0}^1\, \hI_{r,s,b}\, ,
\ee
where
\be\label{enn6rpt}
\hI_{r,s,b} = \int_{\FF} \frac{d^2\tau}{\tau_2} 
\left[\sum_{\stackrel{m_1, n_1\in \zzz, m_2\in 
\zzz/N}{n_2\in N\zzz+ r, 
j\in 2\zzz + b}}
q^{p_L^2/2} \bar q^{ p_R^2/2} e^{2\pi i m_2 s} 
h_b^{(r,s)}(\tau)- \delta_{b,0} \delta_{r,0} c^{(0,s)}_0(0)
\right]\, ,
\quad
q\equiv e^{2\pi i\tau}\, ,
\ee
\bea{e7nrpt}
{1\over 2} p_R^2 &=& {1\over 4 \det Im  \Omega} |-m_1 \rho  +
m_2 + n_1 \sigma + n_2 (\sigma\rho -v^2) + j v |^2, \nonumber \\
{1\over 2} p_L^2
&=& {1\over 2}  p_R^2 + m_1 n_1 + m_2 n_2 + {1\over 4} j^2\, ,
\eea
\be\label{edefomegarpt}
\Omega=\pmatrix{\rho  & v \cr v  & \sigma}\, ,
\ee
and
$\FF$ denotes the fundamental region of $SL(2,\ZZZ)$ in the
upper half plane. $\hI$ is related to $\Phi$ by the relation
\be\label{enn9rpt}
\hI(\rho ,\sigma, v ) = -2 \ln \left[
 ( \det\,{\rm Im}\Omega)^k \right] - 2\ln \Phi (\rho ,\sigma,v )
- 2\ln \overline{{\Phi}(\rho ,\sigma,v )} +\hbox{constant}
\ee
where
\be\label{ekvaluerpt}
k={1\over 2}\, \sum_{s=0}^{N-1} \, c_0^{(0,s)}(0)\, .
\ee

By making the rearrangement
\ben \label{etrs1rpt}
(m_1, n_1, m_2, n_2, j) &\to& (-n_1, -m_1, m_2, n_2, -j),  \nonumber \\
\hbox{and}
\quad 
(m_1, n_1, m_2, n_2, j) &\to& (m_1 - n_1 - j, n_1, m_2, n_2, j+ 2 n_1)\, ,
\een
one can prove the invariance of $\hI$ and of $\Phi$ under the
transformations
\be \label{etrs2rpt}
(\rho,\sigma, v) \to (\sigma, \rho, -v), \qquad \hbox{and}
\qquad
(\rho,\sigma, v) \to (\rho,\sigma+\rho-2v, v-\rho) \, .
\ee
As in the case of $\ZZZ_2$ twisted index in type IIB string theory
on $K3\times T^2$, the symmetry \refb{etrs2rpt} can be
used to prove the S-duality invariance of the partition function.
More generally $\Phi$ transforms as a modular form of weight
$k$ under a subgroup of $O(3,2;\ZZZ)$,
cconsisting of $O(3,2)$ matrices which, acting as in \refb{eso32},
preserves the restrictions on $(\vec m, \vec n, j)$ in the sum in
\refb{enn6rpt}, and preserves $m_2$ mod 1 so that the 
$e^{2\pi i m_2 s}$
factor in \refb{enn6rpt} also remains invariant. The generators of this
subgroup of $Sp(2,\ZZZ)$ have been given explicitly in
\cite{0609109,0708.1270}.

The zeroes and poles of $\Phi$ can also be found from $\hI$
following the analysis of \cite{0605210,0609109} reviewed
in \cite{0708.1270} (appendix D).
The result is that the zeroes and poles of $\Phi$ are of the following
form:
\ben \label{efzp}
&& \Phi \sim \left( n_2 ( \sigma \rho  -v ^2) + jv  + 
n_1 \sigma  -\rho m_1 + m_2
\right)^{\sum_{s=0}^{N-1}
e^{2\pi i m_2 s }\, c_1^{(r,s)}(-1)}\, , \cr
&& m_1, n_1, n_2 \in \ZZZ, \quad j\in 2\ZZZ+1, \quad
m_2\in \ZZZ/N, \quad r=n_2 \, \hbox{mod}\, N, \quad 
m_1 n_1 + m_2 n_2 +\frac{j^2}{4} = {1\over 4}\, . \nonumber \\
\een
For $N\ge 5$, we also have additional zeroes/poles of the type
\ben \label{eszp}
&& \Phi \sim \left( n_2 ( \sigma \rho  -v ^2) + jv  + n_1 \sigma  
-\rho m_1 + m_2
\right)^{\sum_{s=0}^{N-1}
e^{2\pi i m_2 s }\, c_1^{(r,s)}(-1+{4p\over N})}\, , \cr
&& m_1, n_1, n_2 \in \ZZZ, \quad j\in 2\ZZZ+1, \quad
m_2\in \ZZZ/N, \quad r=n_2 \, \hbox{mod}\, N,  \nonumber \\
&& 
m_1 n_1 + m_2 n_2 +\frac{j^2}{4} = {1\over 4}-{p\over N},
\quad p\in\ZZZ, \quad 1\le p< {N\over 4}\, .
\nonumber \\
\een

It was argued in \cite{0708.1270} (appendix D) that the exponents
in \refb{efzp} and \refb{eszp} are always negative 
or zero for $r\ne 0$ mod
$N$, hence they correspond to poles of $\Phi$.\footnote{For
prime values of $N$ explicit computation using known values
of $c^{(r,s)}_1(u)$ shows that the exponent in \refb{efzp}
always vanishes for $r\ne 0$, whereas the exponent in \refb{eszp}
is given by $-48/(N^2-1)$ for $N=5$ and $N=7$ when
$m_2 r=-1/N$ and vanishes otherwise.} Since our main interest is
in determining the poles of the partition function which come from
the zeroes of $\Phi$, we can ignore the contribution from the 
$r\ne 0$ terms.
For $r=0$ we must
have $n_2=0$ mod $N$. In this case
the constraints on $(m_i, n_i, j)$ forces $p$ to vanish
in \refb{eszp}, reducing it to the case described in \refb{efzp}.
Finally using the identities reviewed in \cite{0708.1270}
\ben \label{ideen1}
\MM=K3 &:&  \sum_{s=0}^{N-1}
e^{2\pi i l s / N}\, c_1^{(0,s)}(-1) = \cases{
\hbox{2 for $l\in N\ZZZ$} \cr \hbox{0 otherwise}}\, ,
\nonumber \\
\MM = T^4 &:& \sum_{s=0}^{N-1}
e^{2\pi i l s / N}\, c_1^{(0,s)}(-1) = \cases{
\hbox{2 for $l\in N\ZZZ$} \cr \hbox{$-1$ for
$l\in N\ZZZ\pm 1$} \cr \hbox{0 otherwise}}\, ,
\een
we see that only zeroes of $\Phi$, both for $K3$ and $T^4$,
arise from the choice $m_2\in \ZZZ$ in \refb{efzp} and 
are of the form
\ben \label{efinz}
\Phi &\sim& \left( n_2 ( \sigma \rho  -v ^2) + jv  + n_1 \sigma  
-\rho m_1 + m_2
\right)^2 \nonumber \\
&& m_1, n_1, m_2\in \ZZZ, \quad   n_2\in N\ZZZ, \quad
j\in 2\ZZZ+1, \quad 
m_1 n_1 + m_2 n_2 +\frac{j^2}{4} = {1\over 4}\, .
\een

The knowledge of the zeroes of $\Phi$ gives us two important
informations. First of all the zeroes of the form described in
\refb{epolestr}, obtained by choosing $n_2=0$ in
\refb{efinz}, give us information on wall crossing for
the decay \refb{edecay}. Again using S-duality transformation
all such decays can be related to the decay given in \refb{esimpled}
 with the corresponding wall at $v=0$. Using
 \refb{enn9c} and the identity 
 \be \label{eidentityrpt}
\sum_{b=0}^1\sum_{j\in 2\zzz+b}
 c_b^{(r,s)}(4n - j^2) = \delta_{n,0}\, 
\{c_0^{(r,s)}(0) + 2\, c_1^{(r,s)}(-1)\}\, ,
\ee
 one finds that
 near $v=0$,
 \be \label{enear1}
\Phi(\rho,\sigma,v) 
=
 -4\, \pi^2 \, v^2\,  g(\rho) \, g(\sigma) + \OO(v^4)\, ,
\ee
where
\be \label{egrho2}
g(\rho) = C^2\, e^{2\pi i \wh\alpha\rho}\, 
\prod_{r=0}^{N-1}\,  \prod_{n=1}^\infty
\Big\{ 1 - e^{2\pi i r / N} \, e^{ 2\pi i n\rho 
}
\Big\}^{ \sum_{s=0}^{N-1}
e^{-2\pi i rs/N}  \left(c^{(0,s)}_0(0) + 2 c^{(0,s)}_1(-1)\right)
}\, .
\ee
The constant $C$ has been defined in \refb{enn9d}.
Using \refb{enear1} 
we see that the jump in $B^g_6$ across the
wall of marginal stability, given by the 
residue of the integrand in
\refb{ebg6new} at the pole at $v=0$, is given by
\be \label{erespole}
(-1)^{Q\cdot P+1} \, Q\cdot P \,  f(Q) \, f(P)\, ,
\ee
where
\be \label{efp}
f(Q) = \int_0^1 d \rho \, e^{-i\pi Q^2 \rho} \, (g(\rho))^{-1}
\, d\rho\, .
\ee
It is easy to check, {\it e.g.} by computing the index
$B^g_4$ 
of a Kaluza-Klein monopole carrying momentum along $S^1$,
that $f(Q)$ and $f(P)$ have the interpretation of the index
$B^g_4$ for dyons carrying charge $(Q,0)$ and $(0,P)$
respectively. Thus we get
\be \label{egenwall}
\Delta B^g_6 = (-1)^{Q\cdot P+1} \, (Q\cdot P) \, 
B^g_4((Q,0))\, B^g_4((0,P))\, ,
\ee
in agreement with the expected wall crossing 
formula for the index $B_6$.

The second application of the knowledge of the zeroes
of $\Phi$ is in the determination of the asymptotic behaviour of
$B^g_6$ for large charges. They are controlled by the zeroes of
$\Phi$ given in \refb{efinz} for $|n_2|>0$.
The constraint $n_2\in N\ZZZ$ in \refb{efinz}
implies that the lowest value of 
$|n_2|$ other than zero for which there is a pole is $|n_2|=N$.
The analysis of the asymptotic growth of the index, which is
controlled by this pole, then tells us that the index grows as
\be \label{emicro}
\exp\left(\pi\sqrt{Q^2 P^2 - (Q\cdot P)^2}/N\right)\, .
\ee

\sectiono{Macroscopic Explanation from Quantum Entropy Function}
\label{ss6}

The near horizon attractor geometry of an extremal black hole in
four dimensions contains an $AdS_2\times S^2$ factor. After
euclidean continuation the metric on $AdS_2\times S^2$ takes the form
\be \label{eqe1}
ds^2 = v_1 (d\eta^2 + \sinh^2 \eta d\theta^2) 
+ v_2 (d\psi^2 + \sin^2 \psi d\phi^2)\, ,\quad 0\le\eta<\infty, \quad
\theta\equiv\theta+2\pi, \quad (\psi,\phi)\in S^2\, ,
\ee 
where $v_1$ and $v_2$ are constants whose values are determined by the
charges carried by the black hole. Besides the metric the background
also has non-vanishing fluxes through various cycles and constant
vacuum expectation values of the scalars which we have not written
down explicitly. These background fields respect the $SO(3)\times
SO(2,1)$ isometry of $AdS_2\times S^2$.

According to the quantum entropy function 
proposal\cite{0809.3304,0903.1477} (see also \cite{0809.4264})
the degeneracy
associated with the black hole horizon is given by an appropriate
path integral of string theory in the euclidean
near horizon geometry of the
black hole. 
More precisely the 
degeneracy associated with the black hole horizon is given by
\be \label{ealt}
d_{hor}(\vec q) = Z^{finite}\, ,
\ee
\be \label{ezfinite}
Z \equiv \left\langle \exp[-
i  q_i\ointop d\theta \, A^{(i)}_\theta]
\right\rangle_{AdS_2}\, .
\ee
Here $\langle~\rangle$ denotes the result of path integral over
the string fields, $q_i$ are the electric charges describing fluxes
of the $U(1)$ gauge fields $A_\mu^{(i)}$
in $AdS_2$ and $\ointop d\theta$ denotes
an integral along the boundary of $AdS_2$. The superscript `finite'
refers to the infrared finite part of the amplitude defined as
follows. If we carry out the path integral by putting an infrared
cut-off at $\eta=\eta_0$ so that the boundary has a finite
length $L$, then we may express $Z$ computed via
\refb{ezfinite} as\cite{0809.3304,0903.1477}
\be \label{edeffin}
Z = e^{C L + \OO(L^{-1})} \, Z^{finite}\, .
\ee
for some $L$ independent constant $C$. As indicated in
\refb{edeffin}, $Z^{finite}$ is obtained
by removing the $e^{CL}$ factor from $Z$ and 
taking the $L\to\infty$
limit. This prescription for computing $d_{hor}(\vec q)$
follows naturally from $AdS_2/CFT_1$ correspondence and
reduces to the exponential of the Wald entropy in the
classical limit\cite{0809.3304}.

In \refb{ezfinite} the path integral is to be carried out
over all string field configurations whose asymptotic geometry
coincides with the attractor geometry. In particular 
in integrating over the gauge fields we must
keep the electric fields fixed at infinity and allow the constant
modes of $A^{(i)}_\theta$ to fluctuate\cite{0809.3304}. 
This follows from the
fact that the electric field modes represent non-normalizable 
deformations of $AdS_2$ whereas the constant $A^{(i)}_\theta$ modes
represent normalizable deformations.
With this prescription one can argue, 
via $AdS_2/CFT_1$ correspondence,
that the $d_{hor}$ defined in \refb{ealt}, \refb{ezfinite} measures the
degeneracy of a dual quantum mechanical system whose Hilbert space
is degenerate and
finite dimensional, containing the grounds states of the
black hole carrying a fixed set of charges.

Since $d_{hor}$ measures the degeneracy associated with the horizon
whereas in the microscopic analysis we calculate an index, one might
wonder how we can compare the two quantities. For the helicity trace
index $B_{2n}$
this issue was addresed in \cite{0903.1477} 
where the following explanation
was offered. The main idea is to try to compute the index
$B_{2n}$ on the macroscopic side as well and then compare this
with the microscopic result.
The first step is to note that the 
factors of  $2h$ inserted into the trace
are used in soaking up the
fermion zero modes associated with the 
broken supersymmetries. Typically these modes are always part of
the hair degrees of freedom of the black hole, \i.e.\ live outside
the horizon.
This allows us to relate $B_{2n}$ to $Tr(-1)^{2h}$ 
associated
with the horizon degrees of freedom\cite{0903.1477}. 
In fact if the only hair degrees
of freedom are the fermion zero modes associated with broken
supersymmetry then, up to a sign, 
the macroscopic $B_{2n}$ can be shown to
be equal to $Tr (-1)^{2h}$ associated with the horizon degrees of
freedom. The second step is to note that although from the
point of view of the asymptotic observer $2h$ measures angular
momentum, in $AdS_2$ it can be regarded as an electric
charge of  the $U(1)\subset SU(2)$ 
gauge field $\AAA$  associated with
the rotational isometry of $S^2$\cite{0606244}. 
Thus while carrying out 
the path integral this charge must be
fixed, and in fact has zero value since the attractor geometry is
spherically symmetric.\footnote{In classical supergravity in
four dimensional Minkowski space all supersymmetric black holes
are spherically symmetric. One can argue that 
supersymmetric black holes
whose near horizon geometry has an $AdS_2$ factor must be
spherically symmetric even in the full theory.
For this we note that due to the presence of the $AdS_2$ throat, 
the full symmetry 
algebra at the horizon
contains an $sl(2,R)$ subalgebra besides the 
supersymmetry generators. The (anti-)commutators of the $sl(2,R)$
and the supersymmetry generators then generate the $su(1,1|2)$
algebra which contains $su(2)$ as its subalgebra.}
Thus we have $(-1)^{2h}=1$ and 
$B_{2n}$ can be directly related to $d_{hor}$ 
defined in \refb{ealt}, \refb{ezfinite}.
Put another way, computation of $Tr (-1)^{2h}$ can be expressed
as a path integral of the kind described in \refb{ezfinite}, but
with a twisted boundary condition that requires the fields to
transform by $(-1)^{2h}$ as $\theta$ changes by $2\pi$. This can
be regarded as a Wilson line of $\AAA$ along the boundary circle.
But while carrying out the path integral over the gauge fields we
are instructed to integrate over the Wilson lines of $\AAA$ along the
boundary circle keeping the electric fields fixed. Thus a
background Wilson line along the boundary circle can be
removed by a shift in the integration
variables and does not affect the final value
of the path integral.

If we try to follow a similar logic for the index $B^g_{2n}$ then
the first part of the argument goes through as usual, \i.e.\ the
factors of $(2h)^{2n}$ inserted into the trace are absorbed by the
fermion zero modes living on the hair. At the end we are left
with $Tr(-1)^{2h} g$ associated with the horizon degrees of
freedom. This can be expressed as a path integral similar to the
one described in \refb{ealt}, \refb{ezfinite}, but with a twisted
boundary condition on the fields which require the fields to
transform by $g (-1)^{2h}$ as $\theta$ changes by $2\pi$. 
Let us call this partition function $Z_g$.
Of these
the factor of $(-1)^{2h}$ can be removed by the same argument
described above. For the factor of $g$ there are two possibilities.
If $g$ can be regarded as a rigid gauge transformation
for some U(1) gauge field living on $AdS_2$
 -- like
$(-1)^{2h}$ as an element of the gauge group associated with the 
rotational invariance on $S^2$ -- then the effect of the 
insertion of $g$ is a background value of the Wilson line associated
with the gauge group. Since in carrying out the path integral
we integrate over the mode representing
the Wilson line at the boundary, the effect of $g$ insertion 
has no effect. Alternatively one can say that since the electric charges
associated with all the gauge fields are fixed at the boundary, all
states which contribute to $d_{hor}$ have the same
value of $g$ and hence the effect of insertion of $g$ into the trace
is trivial. On the other hand if $g$ is not an element of a $U(1)$
gauge group then there is no such interpretation. In this case the
attractor geometry is not a valid saddle point in the theory, since
the boundary circle along which we insert the twist by $g$ is
contractible at the center of $AdS_2$ ($\eta=0$) and a twist by
$g$ will produce a singularity at the center of $AdS_2$.

This however is not 
the end of the story. In carrying out the string
path integral we are instructed to integrate over all configurations
preserving the required boundary condition at $\eta\to\infty$.
So we can look for other saddle points.
One of the criteria one must use in searching for these saddle
points is supersymmetry; if the saddle point breaks too much
supersymmetry then integration over the associated fermion zero
modes will make the path integral vanish. 
It was shown in \cite{0905.2686} 
that if we take an orbifold of the original
geometry by a transformation that involves equal amount of
rotation in $AdS_2$ and $S^2$, possibly accompanied by another
symmetry that commutes with supersymmetry, then the resulting
orbifold, if consistent, preserves 
the necessary number of
supersymmetries so that the contribution from this saddle point to
the path integral does not vanish due to integration over the
fermion zero modes. 
This suggests the following procedure for constructing a saddle
point that contributes to $B^g_6$:
we take the original
attractor geometry geometry and then take an orbifold of this by
a $\ZZZ_N$ transformation that combines the action of $g$ with a shift
of $\theta$ by $2\pi/N$ and a shift of $\phi$ by $2\pi/N$.
It can be shown following 
\cite{0810.3472} 
that 
this geometry satisfies 
the required boundary condition as $\eta\to\infty$.
For this we need to carry out a rescaling $\theta\to \theta/N$,
$\eta\to \eta + \ln N$ so that in the new coordinate
system the $AdS_2$ part of the metric takes the form:
\be \label{eads2}
v_1 \left[ d\eta^2 + N^{-2} \sinh^2\left(\eta + \ln N\right)
d\theta^2\right] 
= v_1 \left[ d\eta^2 + \left\{\sinh\eta + {1\over 2} \, e^{-\eta}\,
(1-N^{-2})\right\}^2 
d\theta^2\right]\, .
\ee
Clearly as $\eta\to\infty$  the metric approaches that of
$AdS_2$.
The orbifold group now acts as $\theta\to\theta+2\pi$, $\phi\to
\phi+{2\pi\over N}$ together with an action of $g$. The $g$
action is exactly as required for computing the $g$ twisted
index. On the other hand
the  $2\pi/N$ rotation in $\phi$ is
part of a gauge transformation from the point of view
of the theory on $AdS_2$ and hence, by our previous
argument, has no effect on the
macroscopic computation of the index, except possibly an
overall phase. Put another way, the path integral involves integrating
over all values of the Wilson line at $\infty$, and hence a saddle
point corresponding to any specific value of the Wilson lline
is an admissible configuration in the path integral.
Thus we conclude that this saddle point contributes 
to $Z_g$.

It follows from the analysis of \cite{0810.3472} 
that the semiclassical contribution to $Z^{finite}_g$ from this
saddle point is given by $\exp(S_{wald}/N)$ where $S_{wald}$ is the
Wald entropy of the BPS black hole carrying the same set of charges.
The argument is fairly straightforward: if we keep the cut-off
of $\eta$ fixed as we take the orbifold action, then  
the classical action gets divided by $N$. On the other hand the
length of the boundary also gets divided by $N$. Thus after 
subtracting
the term proportional to the length of the boundary to extract the
finite part of the action, we find that the finite part of the action
for the orbifold is $1/N$ times the finite part of the action for the
original attractor geometry. Since the latter is given by
$S_{wald}=
\pi\sqrt{Q^2 P^2 - (Q\cdot P)^2}$, 
we get
\be \label{ehormac}
Z_g^{finite}\sim \exp\left[
{\pi\over N} \,\sqrt{Q^2 P^2 - (Q\cdot P)^2}\right]\, .
\ee
This is in agreement with the microscopic result given in
\refb{emicro}.

One possible problem with this construction however is that
this saddle point has a $\ZZZ_N$ orbifold singularity on a codimension
eight subspace that lies at the center
of $AdS_2$, either at the north or the south pole of $S^2$ and at the
fixed points of $g$ in $\MM$. In the absence of  flux this is a 
consistent orbifold of string theory, but it is not clear if the 
presence of the background flux
makes the orbifold inconsistent. 
If the attractor geometry
contains a circle $C$ then one way to avoid the existence of the
orbifold singularity is to accompany the $\ZZZ_N$ transformation
also by $1/N$ unit of shift along 
$C$\cite{0903.1477,0904.4253,
0908.0039}.
 The $\ZZZ_N$ now
acts freely on the attractor geometry and hence there is no fixed point.
By our previous argument the $1/N$ unit of shift along $C$ under
$\theta\to\theta+2\pi$ effectively amounts to switching on a
Wilson line at the boundary of $AdS_2$ and hence is an
admissible saddle point.
However in this case there is another subtlety, arising from the fact
that
at the origin of $AdS_2$ \i.e.\ at
$\eta=0$, the shift along $\theta$ is irrelevant. Thus we have an
identification under $\phi\to \phi+2\pi/N$ together with the action
of $g$ and $1/N$ unit of 
shift along $C$.
As a result at the center of $AdS_2$
any magnetic 
flux through the three cycle $S^2\times C$, possibly accompanied
by some other cycles of $\MM\times T^2$, will get divided
by $N$, and unless the original flux through this cycle is a
mutiple of $N$, the
orbifold is not a consistent background of string theory. Thus
we must judiciously choose the circle $C$ such that $C\times S^2$
either does not carry any magnetic 
flux or carries  $N$ units of magnetic flux.
In the present example $S^1$ provides us with such a circle, in
a spirit close to the one discussed in \cite{0904.4253}.

This finishes our proof that the macroscopic `entropy'  associated
with the index $B^g_6$ grows as $\pi\sqrt{Q^2 P^2 - (Q\cdot P)^2}/N$
for large charges, in agreement
with the microscopic results.

\sectiono{Discussion} \label{s7}

In this paper we have computed the twisted helicity trace
index for a class of dyons in type II string theory compactified
on $T^6$ or $K3\times T^2$ and studied their properties.
We have also provided a macroscopic explanation for the
observed asymptotic growth of this index by identifying the
leading saddle point in the path integral
which contributes to this index.

One of the special features of an index of this type is that in
deriving the various general properties of this index we can
focus on supersymmetries which commute with this 
twist.
In particular if the full theory has $M$ supersymmetries,
and if only $N$ of these commute with this twist, then the
general properties of the index ({\it e.g.} under wall crossing)
will be similar to those of the usual helicity trace index in a
theory with $N$ supersymmetries.

We shall now give some more examples of such indices which
could have potential application.
Consider type IIB / IIA
string theory compactified on $T^6$. The spectrum of elementary
strings in this theory contains quarter BPS states obtained by keeping
the right-moving world-sheet degrees of freedom in their
ground state and exciting only the left-moving modes. Since these
states break 24 out of 32 supersymmetries, the relevant index for
these states is $B_{12}$. 
This can be easily computed in the light-cone gauge Green-Schwarz
formulation, where the world-sheet degrees of freedom consist
of 8 scalars, 8 left-moving fermions and eight right-moving fermions. 
However it turns out that due to an
extra cancelation between the bosonic and fermionic states
in the left-moving sector of the world-sheet,
$B_{12}$ grows with charges at a rate
much lower than the rate at which the absolute degeneracies
grow\cite{0507014,0908.3402}. 
Thus the contribution from most of the states cancel
due to the cancelation between the
contributions from fermionic and bosonic states, -- indeed whereas
the degeneracies grow exponentially with the charges, the index
grows only as a power of the charges.

However consider now the subspace of the full moduli space
of the theory where we set all the Ramond-Ramond (RR) moduli
to zero. In this subspace the theory has a discrete symmetry
denoted as $(-1)^{F_L}$ that changes the signs of all the
RR and R-NS sector states. Since elementary string states
only carry charges under the $NS-NS$ sector gauge fields, these
charges are automatically invariant under $(-1)^{F_L}$. Thus
we can choose $g=(-1)^{F_L}$ for defining a new index
$B^g_{2n}$. For elementary string states in the ground state
of the right-moving sector but with arbitrary
excitations in the left-moving sector of the world-sheet
all the 16 supersymmetries in the R-NS sector, and
8 supersymmetries in the NS-R sector are broken. Thus we
have 16 $g$-odd and 8 $g$-even broken supersymmetries, and
these elementary string states contribute to $B^g_4$. 
The eight $g$-invariant fermion zero modes from the right-moving
sector are soaked up by the factor of $(2h)^4$, whereas the
eight fermion zero modes from the left-moving sector are
even under $(-1)^{F_L} (-1)^{2h}=(-1)^{F_R}$ and hence 
gives a factor of $2^4=16$ in the trace. The rest of the computation
involves keeping the right-movers in their ground state and computing
the degeneracy of states created by the left-moving oscillators without
any weight factor.  
As a result there is no
cancelation, and $B^g_4$ grows at the same rate as the degeneracy,
\i.e.\ exponentially, according to the Cardy formula for a CFT with
eight bosons and eight fermions.

The next example involves quarter BPS states in the heterotic string 
theory on $T^6$. Since these states break 12 out of  16 supersymmetries
the appropriate index is $B_6$. But we can consider a subspace
of the moduli space where $T^6$ factorizes into a product of
$T^4$ and $T^2$. In this subspace the theory has an extra discrete
symmetry that involves reversing the sign of the four coordinates
of $T^4$. We identify this as our symmetry $g$.
If we consider charge vectors which carry only momentum
and winding charges, and KK monopole and H-monopole charges
along the two circles of $T^2$ then these charges are invariant under
$g$. Thus we can define an index $B^g_{2n}$ for these charges.
Under the action of $g$ half of the 16 supersymmetries are odd
and half are even, but one can show that all the 8 $g$-odd
supersymmetries are
broken by the dyon. Thus these dyons have 4 broken supersymmetries
which are even under $g$, and hence contribute to the index
$B^g_2$. We expect this index to have properties similar to that of
$B_2$, -- the index that is relevant for capturing the spectrum of
half BPS states in $\NN=2$ supersymmetric string 
theories.\footnote{If instead we take an orbifold
of the theory by $g$ we may get an $\NN=2$ theory, similar
to the S-T-U model analyzed in \cite{0711.1971}. We expect that
the analysis of $B^g_2$ in the $\NN=4$ theory may be simpler
than the one for  the S-T-U model.}
In
particular the wall crossing formula for this index will be controlled
by the Kontsevich-Soibelman formula\cite{kont,0807.4723}.

\medskip

\noindent {\bf Acknowledgment:} I would like to thank 
Atish Dabholkar, Justin David, Joao Gomes, Dileep Jatkar and
Sameer Murthy for useful discussions and Boris Pioline for useful
comments on the manuscript.  This work was supported
in part by the JC Bose fellowship of the Department of Science and
Technology, India and the Blaise Pascal Chair, France.

%\small
%\baselineskip 16pt

\end{document}